\documentstyle[epsf]{article}
\title{The spin-$\frac{1}{2}$ transverse $XX$ chain
       \protect\\
       with regularly alternating bonds and fields}

\author{Oleg Derzhko$^{\dagger,\ddagger}$,
        Johannes Richter$^{\star}$ 
        and Oles' Zaburannyi$^{\ddagger}$\\
\small {\em {$^{\dagger}$Institute for Condensed Matter Physics,}}\\
\small {\em {1 Svientsitskii St., L'viv--11, 290011, Ukraine}}\\
\small {\em {$^{\ddagger}$Chair of Theoretical Physics, 
             Ivan Franko State University of L'viv,}}\\
\small {\em {12 Drahomanov St., L'viv--5, 290005, Ukraine}}\\
\small {\em {$^{\star}$Institut f\"{u}r Theoretische Physik,
             Universit\"{a}t Magdeburg,}}\\
\small {\em {P.O. Box 4120, D--39016 Magdeburg, Germany}}}

\date{April 29, 1999}

\begin{document}

\maketitle

\begin{abstract}
We use continued fractions for a study of the thermodynamic properties 
of the periodic nonuniform spin-$\frac{1}{2}$ isotropic $XY$ chain
in a non--random/random (Lorentzian) transverse field.
The obtained results permit to examine the influence of
a magnetic field
and randomness
on the spin--Peierls dimerization. 
\end{abstract}

\vspace{1cm}

\noindent
{\bf PACS numbers:}
75.10.-b

\vspace{1cm}

\noindent
{\bf Keywords:}
Spin-$\frac{1}{2}$ $XY$ chain;
Periodic nonuniformity;
Lorentzian disorder;
Density of states;
Magnetization;
Spin--Peierls dimerization\\

\vspace{1mm}

\noindent
{\bf Postal address:}\\
{\em
Dr. Oleg Derzhko (corresponding author)\\
Institute for Condensed Matter Physics,\\
1 Svientsitskii St., L'viv-11, 290011, Ukraine\\
Tel: (0322) 42 74 39\\
Fax: (0322) 76 19 78\\
E-mail: derzhko@icmp.lviv.ua

\clearpage

\renewcommand\baselinestretch {2.1}
\large\normalsize

Spin-$\frac{1}{2}$ $XY$ chains provide an excellent ground for a rigorous 
study of different properties of low--dimensional quantum magnetic
systems.
By means of the Jordan--Wigner transformation such spin
models can be mapped onto noninteracting spinless fermions and
many statistical mechanics calculations can be performed exactly.$^1$ 
In the present Letter we shall exploit the spin-$\frac{1}{2}$
isotropic $XY$ 
chain in a transverse field 
(the transverse $XX$ chain) 
for an analysis of the 
effects of periodic nonuniformity on the thermodynamic properties. 

Starting from the late 60s a number of authors considered 
the spin-$\frac{1}{2}$ 
$XY$ chain in a transverse field on two sublattices.$^{2,3}$
Later
it was found that
the $XX$ chain (without transverse field)
represents a simple system with
a spin--Peierls instability.$^4$
Some further studies of the spin--Peierls instability 
in the spin-$\frac{1}{2}$ $XY$ chains can be
found in Refs. 5, 6
(within the adiabatic limit)
and Ref. 7
(within the non--adiabatic limit).
In what follows
we want to emphasize that the thermodynamic properties of the periodic 
nonuniform transverse $XX$ chains can be examined in a very general fashion 
including into consideration even regularly alternating randomness.
Such an approach is based on the continued--fraction 
representation of the 
(random--averaged)
diagonal one--fermion Green functions. The 
involved continued fractions
can be calculated exactly
for any finite period of regular nonuniformity,
that 
immediately yields 
rigorous explicit expressions for the diagonal one--fermion Green
functions, the density of magnon states, and hence for the thermodynamic 
quantities.
We show that the periodic nonuniformity may induce a number of
interesting
properties,
e.g. plateaus in the low--temperature magnetization
vs field dependence or
a nonzero magnetization in the zero average field.
As an example we 
consider in some detail 
the chain with the random Lorentzian 
transverse field in the case of the period 2.
The obtained density of states for such a chain permits us to
extend the consideration of the influence of an
external field
on
the stability of
the spin--Peierls dimerized phase$^5$ discussing in more
detail
the non--random case and reporting for the first time the effects 
of randomness. The interest in the properties of 
the spin--Peierls systems considerably increased since the discovery of the 
inorganic spin--Peierls compound CuGeO$_3$.$^{8,9}$
The behaviour of the 
spin--Peierls systems in the presence of
an external field or randomness
attracts much interest both from 
the experimental and theoretical 
viewpoints.$^{9}$ We believe that the
discussed possibility to
analyse rigorously the
influence of non--random/random field 
within the frames of the simple model 
may be of interest for the
theory of spin--Peierls compounds.
 
We consider $N\to\infty$ spins $\frac{1}{2}$ on a circle with the 
Hamiltonian of the nonuniform transverse $XX$ model
\begin{eqnarray}
\label{001}
H=\sum_{n=1}^{N}\Omega_ns_n^z
+2\sum_{n=1}^{N}I_n\left(s^x_ns^x_{n+1}+s^y_ns^y_{n+1}\right)
\end{eqnarray}
where the transverse fields
$\Omega_n$
are assumed to be independent
Lorentzian random variables each with the probability distribution
$p(\Omega_n)=\frac{1}{\pi}
\frac{\Gamma_n}
{\left(\Omega_n-\Omega_{0n}\right)^2+\Gamma_n^2}.$
We assume {\em regular} nonuniformity, i.e. the mean value of the
transverse field 
$\Omega_{0n}$
at site $n$, 
the width of its distribution
$\Gamma_n$,
as well as the exchange coupling $I_n$
between neighbouring sites $n$ and $n+1$
vary regularly
from site to site with a period $p$,
i.e. the sequence of parameters is
$\Omega_{01}\Gamma_1I_1
\Omega_{02}\Gamma_2I_2\ldots
\Omega_{0p}\Gamma_pI_p
\Omega_{01}\Gamma_1I_1
\Omega_{02}\Gamma_2I_2
\ldots\Omega_{0p}\Gamma_pI_p\ldots\;.$

Our task is to examine the thermodynamic quantities of
the spin model
(\ref{001}).
Making use of the Jordan--Wigner transformation$^1$
one gets a tight--binding model of spinless fermions on a circle.
Introducing the temperature double--time one--fermion Green functions
and averaging a set of equations for the Green functions
(because of the Lorentzian probability distribution this can be done 
exactly$^{10}$)
one can further make use of the continued--fraction representation for the 
diagonal Green functions exploited earlier for similar models by a number of 
authors.$^{11-14}$
For any finite period of varying the mean value 
of the transverse field, the width of its distribution 
and the exchange coupling 
the 
involved continued 
fractions are periodic and can be evaluated by  
solving quadratic equations.
Thus one gets rigorously the 
diagonal Green functions and therefore the density of magnon  
states $\overline{\rho(E)}$
(a bar denotes the random--averaged quantity), 
which yields the thermodynamic quantities for
the spin model (\ref{001}).
Some examples of such calculations can be found in Ref. 15.
Let us present the result for the density of states 
of a chain having period 2 that will be used afterwards
\begin{eqnarray}
\label{002}
\overline{\rho(E)}
=\frac{1}{2\pi}
\frac{\vert{\cal{Y}}(E)\vert}
{{\cal{B}}(E)},
\nonumber\\
{\cal{Y}}(E)
=(\Gamma_1+\Gamma_2)
\sqrt{\frac{{\cal{B}}(E)+{\cal{B}}^{\prime}(E)}{2}}
-{\mbox{sgn}}\left({\cal{B}}^{\prime\prime}(E)\right)\;
(2E-\Omega_{01}-\Omega_{02})
\sqrt{\frac{{\cal{B}}(E)-{\cal{B}}^{\prime}(E)}{2}},
\nonumber\\
{\cal{B}}(E)
=\sqrt{
\left({\cal{B}}^{\prime}(E)\right)^2
+\left({\cal{B}}^{\prime\prime}(E)\right)^2},
\nonumber\\
{\cal{B}}^{\prime}(E)
=\left[(E-\Omega_{01})(E-\Omega_{02})
-\Gamma_1\Gamma_2-I_1^2-I_2^2\right]^2
-\left[(E-\Omega_{01})\Gamma_2+
(E-\Omega_{02})\Gamma_1\right]^2
-4I_1^2I_2^2,
\nonumber\\
{\cal{B}}^{\prime\prime}(E)
=2\left[(E-\Omega_{01})(E-\Omega_{02})
-\Gamma_1\Gamma_2-I_1^2-I_2^2\right]
\left[(E-\Omega_{01})\Gamma_2+
(E-\Omega_{02})\Gamma_1\right].
\end{eqnarray}
In the non--random case $\Gamma_1=\Gamma_2=0$ Eq. (\ref{002}) 
yields a nonzero value of the density of states 
$\rho(E)=\frac{1}{2\pi}
\frac{\vert 2E-\Omega_{01}-\Omega_{02}\vert}
{\sqrt{-{\cal{B}}^{\prime}(E)}}$
only for 
$-{\cal{B}}^{\prime}(E)
=4I_1^2I_2^2-\left[(E-\Omega_{01})(E-\Omega_{02})-I_1^2-I_2^2\right]^2
>0$
(two subband magnon energy spectrum).
In the uniform case 
$\Omega_{01}=\Omega_{02}=\Omega_0$,
$\Gamma_1=\Gamma_2=\Gamma$,
$I_1=I_2=I$ 
Eq. (\ref{002}) may be rewritten as 
$\overline{\rho(E)}
=\mp\frac{1}{\pi}{\mbox{Im}}\frac{1}
{\sqrt{(E-\Omega_0\pm{\mbox{i}}\Gamma)^2-4I^2}}$ 
that coincides with the result reported in Ref. 16.

Let us briefly discuss some special properties of spin model 
(\ref{001}) induced by regular nonuniformity. Consider at first the 
non--random case $\Gamma_n=0$.
The main result of introducing the periodic nonuniformity is a splitting
of the initial magnon band into several subbands.
The number of subbands does not exceed the period of the nonuniformity.
The splitting of the magnon band into subbands  
has several remarkable consequences for the thermodynamic
properties. For example, the dependence of the transverse magnetization 
$m_z
=-\frac{1}{2}\int\limits_{-\infty}^{\infty}{\mbox{d}}E
\rho(E)\tanh\frac{E}{2kT}$ 
on the transverse field $\Omega_0$
($\Omega_{0n}=\Omega_0+\Omega_{0n}^{\prime},$
$\Omega_{0n}^{\prime}$ are fixed)
at low temperatures is composed of sharply increasing parts separated by 
plateaus, the number of which is determined by the number of subbands.
This can be nicely seen in Figs. 1a, 1b for the case of a chain with 
$p=4$.

One of the interesting magnetic properties of the periodic nonuniform 
spin-$\frac{1}{2}$
transverse $XX$ chain is a possibility of a nonzero transverse
magnetization
$m_z$ at 
zero average transverse field $\sum\limits_{n=1}^N\Omega_{0n}=0.$
A similar property
was found for the spin-$\frac{1}{2}$
transverse $XX$ chain with correlated disorder.$^{17}$
To illustrate this let us consider a chain having the period 4 in 
which the site $n+1$ with the transverse field
$\Omega^{\prime}_{02}<0$ is surrounded by the strong couplings
$I_1=I_2,$ whereas the site $n+3$ with the transverse field
$-\Omega^{\prime}_{02}>0$ is surrounded by the weak couplings
$I_3=I_4=0.$
The transverse fields at site $n$ and site $n+2$
are assumed to be zero, i.e.
$\Omega^{\prime}_{01}=\Omega^{\prime}_{03}=0.$
Naively reasoning one may
expect that
the local transverse magnetization at site $n+1$
has smaller value and opposite direction with respect to
the local transverse magnetization at site $n+3$
and therefore
a nonzero
total transverse magnetization at zero
average transverse field may be anticipated.
The explicit expression for the density of
states for such a chain reads
$\rho(E)
=\frac{1}{4}
\left[
\delta\left(E-\frac{1}{2}(\Omega_{02}-\sqrt{\Omega_{02}^2+8I_1^2})\right)
+\delta\left(E\right)
+\delta\left(E-\frac{1}{2}(\Omega_{02}+\sqrt{\Omega_{02}^2+8I_1^2})\right)
+\delta\left(E+\Omega_{02}\right)
\right]$
(solid curve in Fig. 1a)
and for $T=0$ one gets $m_z=-\frac{1}{8}\ne 0$
at $\sum\limits_{n=1}^N\Omega_{0n}=0$
(solid curves in Figs. 1b, 1c).
The picture becomes more complicated if the weak
couplings $I_3=I_4$ have small but nonzero value (see Fig. 1). In that 
case 
for $T=0$
we have
$m_z=0$ at $\sum\limits_{n=1}^N\Omega_{0n}=0$, 
however, the position of the magnon subbands
provides an interesting dependence of $m_z$ on temperature 
(dashed curve in Fig. 1c).
The latter dependence reminds the `order from disorder'
phenomenon,$^{18}$
i.e. increasing of order with increasing temperature.

The main effect of the diagonal Lorentzian disorder 
is a smearing out of the 
band structure.
The details of the smoothed magnon subbands 
may be essentially different in the case of the uniform disorder 
(all $\Gamma_n$ are the same) 
and the nonuniform disorder 
($\Gamma_n$ are different).
In the latter case one may observe a 
different degree of  
smearing out of the different subbands that results 
in a different degree of smoothing of the different `steps' in the 
step--like dependence of
$\overline{m_z}$ on $\Omega_0$.

Finally, let us discuss
the spin--Peierls dimerization in the presence of 
non--random/random field
following Ref. 4. 
For this purpose we introduce the dimerization parameter 
$\delta,$
$0\le\delta\le 1$
assuming that
$\vert I_1\vert=\vert I\vert(1+\delta),$
$\vert I_2\vert=\vert I\vert(1-\delta).$
The magnetic ground state energy per site
immediately follows from the derived density of 
states (\ref{002}) 
$\overline{e_0(\delta)}
=-\frac{1}{2}\int\limits_{-\infty}^{\infty}{\mbox{d}}E
\overline{\rho(E)}\vert E\vert,$
the elastic energy per site is $\alpha\delta^2$
and hence the total energy is
$\overline{{\cal{E}}(\delta)}
=\overline{e_0(\delta)}+\alpha\delta^2$. 
We shall study the change 
of the total energy 
$\overline{{\cal{E}}(\delta)}
-\overline{{\cal{E}}(0)}$
as a function of the dimerization parameter $\delta$ 
revealing in such a way the instability of the chain with respect to 
dimerization. 
For the non--random case $e_0(\delta)$ can be 
expressed through the elliptic integral of the second kind 
${\mbox{E}}(\psi,a^2)\equiv\int\limits_0^{\psi}{\mbox{d}}\varphi
\sqrt{1-a^2\sin^2\varphi},$
namely,
\begin{eqnarray}
\label{003}
e_0(\delta)
=-\frac{\sqrt{(\Omega_{01}-\Omega_{02})^2+16 I^2}}
{2\pi}
{\mbox{E}}
\left(\psi, \frac{16 I^2(1-\delta^2)}
{(\Omega_{01}-\Omega_{02})^2+16 I^2}\right)
-\vert\Omega_{01}+\Omega_{02}\vert
\left(\frac{1}{4}-\frac{\psi}{2\pi}\right),
\nonumber\\
\psi=
\left\{
\begin{array}{ll}
0,
& {\mbox{if}}\;\;\;
\sqrt{(\Omega_{01}-\Omega_{02})^2+16 I^2}
\le\vert\Omega_{01}+\Omega_{02}\vert,\\
{\mbox{arcsin}}
\sqrt{\frac{4 I^2-\Omega_{01}\Omega_{02}}
{4 I^2(1-\delta^2)}},
& {\mbox{if}}\;\;\;
\sqrt{(\Omega_{01}-\Omega_{02})^2+16 I^2\delta^2}
\le\vert\Omega_{01}+\Omega_{02}\vert
<\sqrt{(\Omega_{01}-\Omega_{02})^2+16 I^2},\\
\frac{\pi}{2},
& {\mbox{if}}\;\;\;
\vert\Omega_{01}+\Omega_{02}\vert
< \sqrt{(\Omega_{01}-\Omega_{02})^2+16 I^2\delta^2}.
\end{array}
\right.
\end{eqnarray}
Eq. (\ref{003}) in the uniform limit 
$\Omega_{01}=\Omega_{02}=\Omega_0$ yields the result 
obtained in Ref. 5. 
We shall also consider the solution $\delta^{\star}$ of the equation 
$\frac{\partial\overline{{\cal{E}}(\delta)}}{\partial\delta}=0$ 
that may 
have a nonzero solution $\delta^{\star}\ne 0.$
In the non--random case the latter solution  
may arise as a solution of the equation 
\begin{eqnarray}
\label{004}
\alpha
=\frac{\sqrt{(\Omega_{01}-\Omega_{02})^2+16 I^2}}
{4\pi(1-\delta^2)}
\left(
{\mbox{F}}
\left(\psi, \frac{16 I^2(1-\delta^2)}
{(\Omega_{01}-\Omega_{02})^2+16 I^2}\right)
-{\mbox{E}}
\left(\psi, \frac{16 I^2(1-\delta^2)}
{(\Omega_{01}-\Omega_{02})^2+16 I^2}\right)
\right)
\end{eqnarray}
where 
${\mbox{F}}(\psi,a^2)\equiv\int\limits_0^{\psi}{\mbox{d}}\varphi
\frac{1}{\sqrt{1-a^2\sin^2\varphi}}$
is the elliptic integral of the first kind.

Let us discuss the
dependence 
$\overline{{\cal{E}}(\delta)}
-\overline{{\cal{E}}(0)}$ 
vs $\delta$ which
manifests 
the spin--Peierls dimerization.
We restrict our discussion to the
uniform case 
$\Omega_{01}=\Omega_{02}=\Omega_0,$
$\Gamma_1=\Gamma_2=\Gamma.$
Consider at first the 
non--random case.
Note that for strong transverse fields
$\vert\Omega_0\vert\ge 2\vert I\vert$ 
from Eq. (\ref{003}) it immediately follows 
that 
${\cal{E}}(\delta)
=-\frac{1}{2}\vert\Omega_0\vert+\alpha\delta^2$ and 
hence ${\cal{E}}(\delta)-{\cal{E}}(0)$ exhibits
a minimum only at
$\delta^{\star}=0$.
This tells us that in strong enough field the uniformed state is
energetically favoured over the dimerized one. Consider further 
weaker fields 
$\vert\Omega_0\vert< 2\vert I\vert$. 
From Eq. (\ref{003}) it follows that 
${\cal{E}}(\delta)$ is influenced by the field for 
$0\le\delta< \frac{\vert\Omega_0\vert}{2\vert I\vert}$ 
but ${\cal{E}}(\delta)$
does not feel the presence of the field for
$\frac{\vert\Omega_0\vert}{2\vert I\vert}\le\delta\le 1$.
On the other hand one can calculate the r.h.s. of Eq. (\ref{004}) 
varying $\delta$ from 0 to 1 finding in such a way the values of 
the dimerization parameter at which the total energy for a certain lattice 
(characterized by $\alpha$)
exhibits an extremum.
As it follows from Eq. (\ref{004}) 
$\frac{\alpha}{2\vert I\vert}\rightarrow\frac{1}{8}$ 
as $\delta\rightarrow 1$ 
that means that for $\frac{\alpha}{2\vert I\vert} < \frac{1}{8}$
the total energy
${\cal{E}}(\delta)$
does not have the extremum
of interest
for
$0\le\delta\le 1$.
Hence, we shall exclude the
soft lattices having small $\alpha$ from further consideration. The 
dependence 
${\cal{E}}(\delta)-{\cal{E}}(0)$ vs $\delta$ at 
different $\Omega_0$ is shown in Fig. 2a. One finds that at $\Omega_0=0$ 
the quantity
${\cal{E}}(\delta)-{\cal{E}}(0)$ exhibits a minimum at 
$\delta^{\star}\ne 0$. The position of this minimum does not change
switching on the field.
However, after a certain value of the transverse field
(for which Eq. (\ref{004}) has
an additional solution
$\delta^{\star}=0$)
an additional local minimum in
${\cal{E}}(\delta)-{\cal{E}}(0)$
appears
at $\delta^{\star}=0$.
The two minima are separated by a maximum
at an intermediate  value of the dimerization parameter.
With further increasing
of $\Omega_0$ the depths of the minima become the same
at a certain value of the transverse field
and then the minimum at $\delta^{\star}=0$
becomes the global one. Finally at a certain field 
(for which Eq. (\ref{004}) has the solution 
$\delta^{\star}=\frac{\vert\Omega_0\vert}{2\vert I\vert}$)
the minimum at the nonzero value of the dimerization parameter
abruptly disappears
that means a complete suppression of 
the dimerization by the field.
The above sketched picture represents a typical scenario
of a first order phase transition
and is illustrated by
the phase
diagram in the $\Omega_0$ --- $\alpha$ plane shown in Fig. 3a.
 
The dependences of
$\overline{{\cal{E}}(\delta)}
-\overline{{\cal{E}}(0)}$ 
vs $\delta$ in the presence of randomness at 
$\Omega_0=0$ and $\Omega_0\ne 0$ are shown in Figs. 2b, 2c. 
In the case $\Omega_0=0$ with increasing of $\Gamma$
the position and the depth of the minimum at 
$\delta^{\star}\ne 0$ continuosly decrease
and for a certain value of the strength of disorder the minimum
occurs already at $\delta^{\star}=0$.
That mean, that a random
field with zero mean value
can suppress the dimerization, too.
However, the dimerization parameter
$\delta^{\star}$ vanishes according to a second order
phase transition scenario in contrast to the transition in a
non--random
field.
The influence of the randomness on the dependence
$\overline{{\cal{E}}(\delta)}
-\overline{{\cal{E}}(0)}$ 
vs $\delta$ at $\Omega_0\ne 0$ can be traced in Fig. 2c. 
The phase diagrams in the plane $\Gamma$ --- $\alpha$ are shown in Figs. 3b, 
3c.

It is necessary to mention that we have considered here
only the stability of the dimerized phase against the uniformed one.
Though, the qualitative picture of a first order phase transition
in a uniform field may be correct, we notice that
both experiments and approximate analytical treatments as well as
exact numerical computations
(mainly for CuGeO$_3$)
show a transition from the dimerized to incommensurate phase.$^{9}$
Evidently, assuming the simple ansatz for the lattice distortion
$\delta_1\delta_2\delta_1\delta_2\ldots\;$,
$\delta_1+\delta_2=0$
we were able to compare the ground state energies only for the dimerized
and uniformed phases.
To detect a transition from the dimerized phase
to the incommensurate phase with increasing of the field
one must analyze the ground state energy of
a chain having sufficiently large period
(say, a chain with $p=12$, for which one can still easily
obtain the density of states$^{19}$).
In the presence of random fields (or/and
random 
exchange couplings)
even more complicated lattice distortions should be examined
and the elaborated approach for 
the calculation of the ground state energy of 
the nonuniform spin-$\frac{1}{2}$ transverse $XX$ chain provides
some
possibilities to perform such an analysis.

To summarize,
we have suggested a general scheme for the study of thermodynamics of 
the regularly 
alternating spin-$\frac{1}{2}$ 
transverse $XX$ chains. It is based on the 
continued--fraction representation for the diagonal one--fermion  
Green functions. We have discussed 
briefly some magnetic properties of 
the periodic nonuniform spin-$\frac{1}{2}$ transverse $XX$ chains such as 
plateaus in magnetization curves or a nonzero magnetization at the zero 
average field, as well as the influence of (Lorentzian) randomness in the 
transverse field on these properties. We have examined 
also the stability of spin--Peierls
dimerized phase in the spin-$\frac{1}{2}$ $XX$ chain in the presence of
non--random/random (Lorentzian) transverse field using the rigorous 
expression for the (averaged over randomness)
ground state energy of the
corresponding alternating chain. 

\vspace{5mm}

The present study was partly supported
by the DFG (projects 436 UKR 17/20/98 and Ri 615/6-1).
It was presented at 
the International Workshop and Seminar on 
Cooperative Phenomena in Statistical Physics:
Theory and Applications (Dresden, 1999).
O. D. is grateful to the
Max--Planck--Institut
f\"{u}r Physik komplexer Systeme
for the hospitality.
He is also indebted to
Mrs. Olga Syska for continuous financial support.

\clearpage

{\large{\bf List of figure captures}}
\normalsize

\vspace{15mm}

FIGURE 1.
Magnetic properties of the chain having period 4.
The local transverse fields are
$\Omega_{0n}=\Omega_0+\Omega^{\prime}_{0n}$,
$\Omega^{\prime}_{01}=\Omega^{\prime}_{03}=0,$
$\Omega^{\prime}_{02}=-\Omega^{\prime}_{04}=-1,$
the exchange couplings are
$\vert I_1\vert=\vert I_2\vert=0.5,$
$\vert I_3\vert=\vert I_4\vert=0$ (solid curves),
$\vert I_3\vert=\vert I_4\vert=0.05$ (dashed curves),
$\vert I_3\vert=\vert I_4\vert=0.25$ (dashed--dotted curves),
$\vert I_3\vert=\vert I_4\vert=0.5$ (dotted curves).
a) Density of states.
b) Transverse magnetization vs transverse field
$\Omega_0$ at $T=0$.
c) Transverse magnetization vs temperature at
$\Omega_0=0$.

\vspace{15mm}

FIGURE 2.
$\overline{{\cal{E}}(\delta)}
-\overline{{\cal{E}}(0)}$
vs $\delta$ for a chain with $\vert I\vert=0.5$, $\alpha=0.5$.
a) Non--random case $\Gamma=0$,
$\Omega_0=0,\;0.01,\;0.02,\;
0.03,\;0.04,\;0.05,\;0.06,\;
0.07,\;0.08,\;0.09,\;0.1$
(from bottom to top).
b) $\Omega_0=0$,
$\Gamma=0,\;
$
$
0.001,\;0.003,\;0.006,\;0.01,\;0.02,\;0.03,\;0.05$
(from bottom to top).
c) $\Omega_0=0.04$,
$\Gamma=0,\;0.001,\;0.003,\;0.006,\;
$
$
0.01,\;0.02,\;0.03,\;0.05$
(from bottom to top).

\vspace{15mm}

FIGURE 3.
Phase diagram of the spin-$\frac{1}{2}$ transverse $XX$ chain
with $\vert I\vert =0.5$
as it follows from the dimerization ansatz
for the ground state energy.
a)  Non--random case $\Gamma=0$.
A --- dimerized state. 
B$_1$, B$_2$ --- both dimerized and uniformed states are possible,
moreover,
in B$_1$ the former one is favourable,
whereas
in B$_2$ the latter one is favourable;
on the line that separates B$_1$ and B$_2$
the depth of both minima in
${\cal{E}}(\delta)-{\cal{E}}(0)$
is equal.
C --- uniformed state.
b) Random case with mean value of the field $\Omega_0=0$.
A --- dimerized state. 
C --- uniformed state.
c) Random case with mean value of the field $\Omega_0=0.04$.
A,
B$_1$,
B$_2$,
C are of the same meaning as in a).

\clearpage

\begin{figure}
\vspace{0mm}
\epsfxsize=200mm
\epsfbox{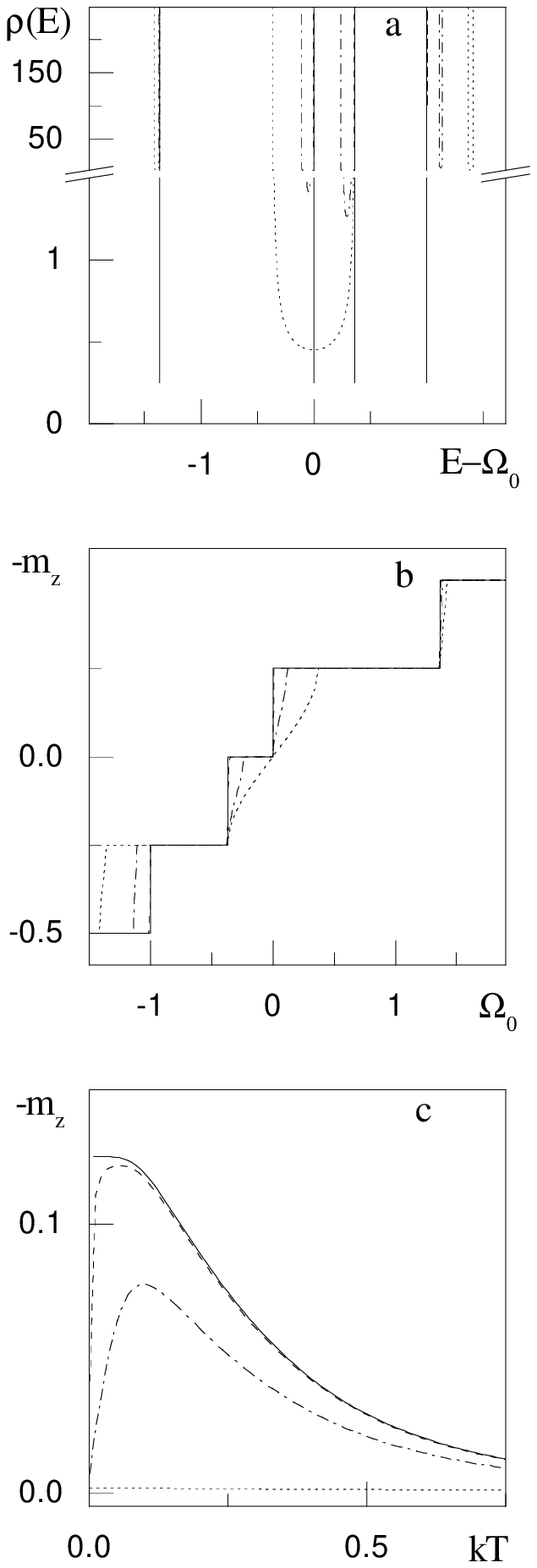}
\vspace{0mm}
\caption{\small{}}
\end{figure}

\clearpage

\begin{figure}
\vspace{0mm}
\epsfxsize=200mm
\epsfbox{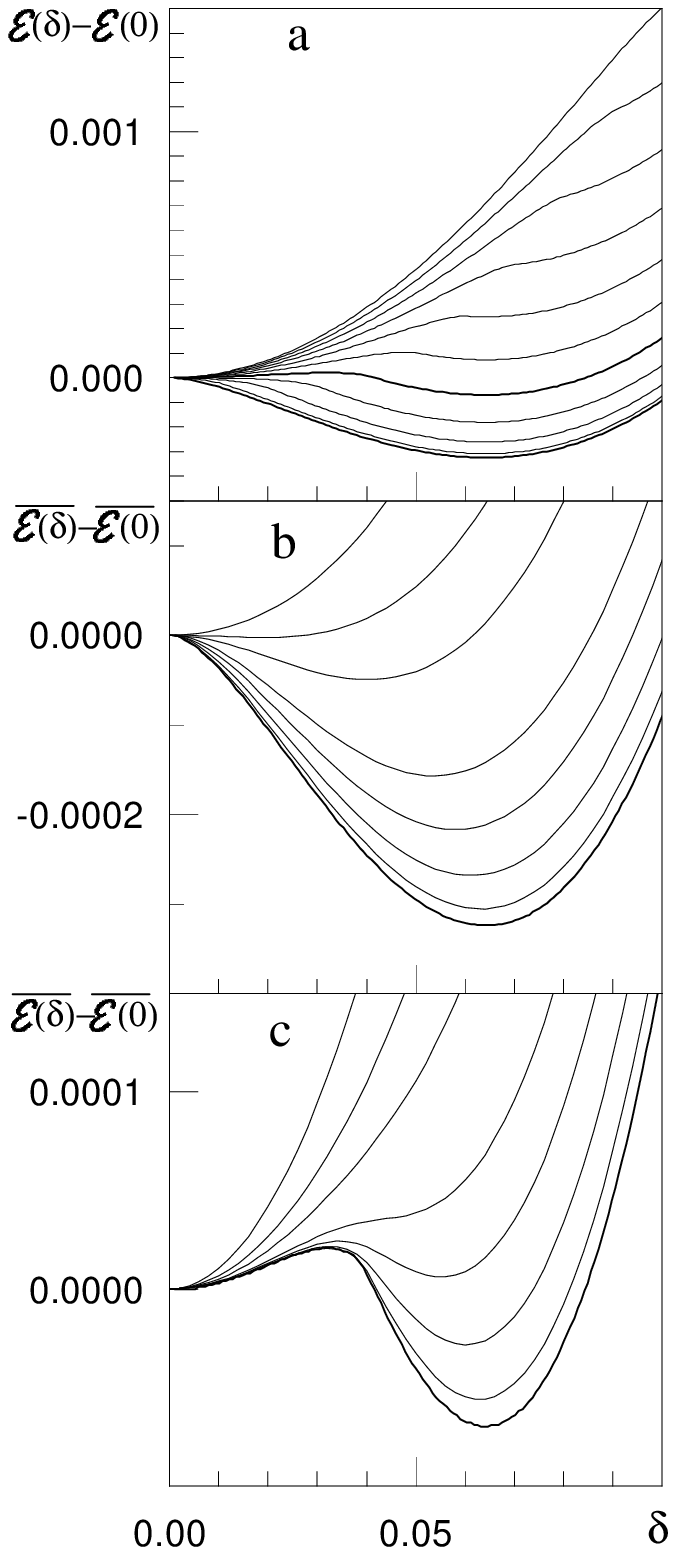}
\vspace{0mm}
\caption{\small{}}
\end{figure}

\clearpage

\begin{figure}
\vspace{0mm}
\epsfxsize=200mm
\epsfbox{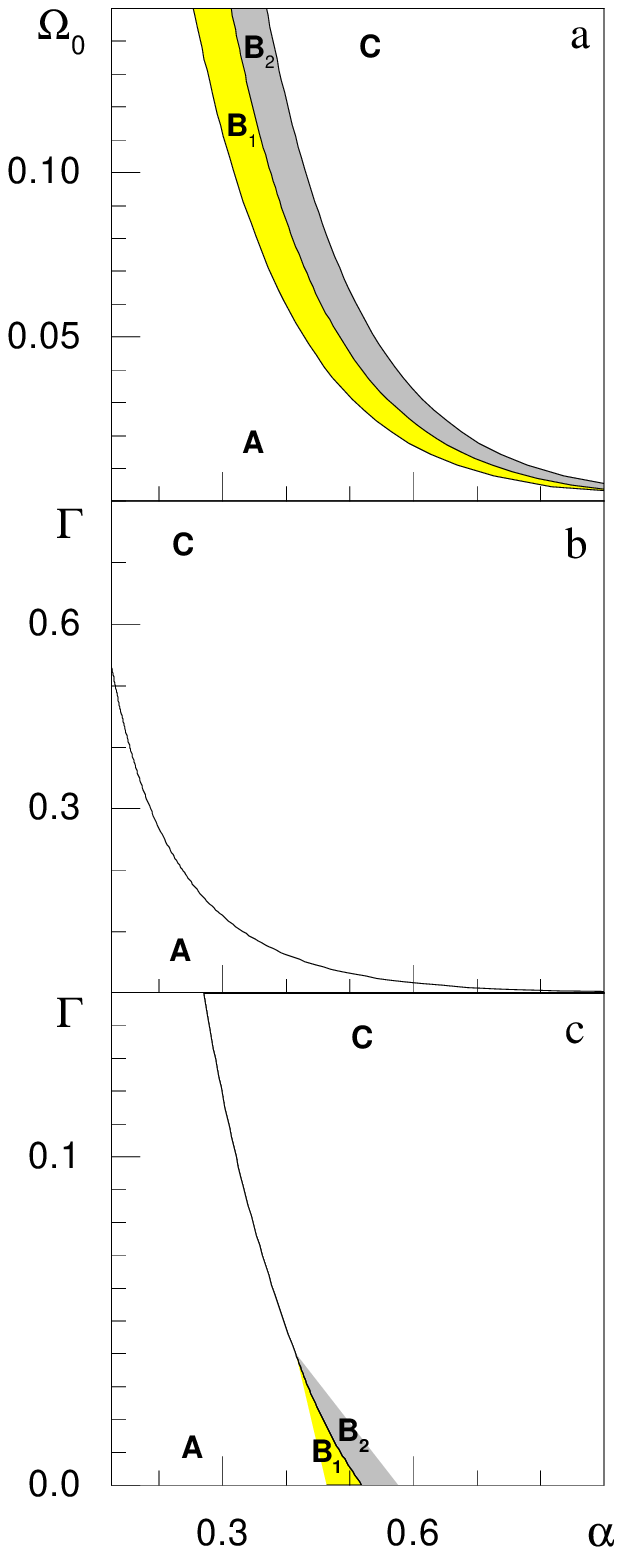}
\vspace{0mm}
\caption{\small{}}
\end{figure}


\begin{thebibliography}{99}
\bibitem{001}
E. Lieb, T. Schultz, and D. Mattis,
Ann. Phys. (N.Y.) {\bf 16,} 407 (1961).
\bibitem{002}
V. M. Kontorovich and V. M. Tsukernik,
Sov. Phys. JETP {\bf 26,} 687 (1968).
\bibitem{003}
J. H. H. Perk, H. W. Capel, M. J. Zuilhof, and Th. J. Siskens,
Physica A {\bf 81,} 319 (1975).
\bibitem{004}
P. Pincus,
Solid State Commun. {\bf 9,} 1971 (1971).
\bibitem{005}
J. H. Taylor and G. M\"{u}ller,
Physica A {\bf 130,} 1 (1985) (and references therein).
\bibitem{006}
K. Okamoto and K. Yasumura,
J. Phys. Soc. Jap. {\bf 59,} 993 (1990) (and references therein);\\
K. Okamoto,
J. Phys. Soc. Jpn. {\bf 59,} 4286 (1990);\\
K. Okamoto,
Solid State Commun. {\bf 83,} 1039 (1992).
\bibitem{007}
S. Sil,
J. Phys.: Condens. Matter {\bf 10,} 8851 (1998).
\bibitem{008}
M. Hase, I. Terasaki, and K. Uchinokura,
Phys. Rev. Lett. {\bf 70,} 3651 (1993).
\bibitem{009}
For a review on CuGeO$_3$ see:
J. P. Boucher and L. P. Regnault,
J. Phys. I France {\bf 6,} 1939 (1996).
\bibitem{010}
P. Lloyd,
J. Phys. C {\bf 2,} 1717 (1969).
\bibitem{011}
S. W. Lovesey,
J. Phys. C {\bf 21,} 2805 (1988).
\bibitem{012}
Ch. J. Lantwin and B. Stewart,
J. Phys. A {\bf 24,} 699 (1991).
\bibitem{013}
J. K. Freericks and L. M. Falicov,
Phys. Rev. B {\bf 41,} 2163 (1990).
\bibitem{014}
R. \L y\.{z}wa, 
Physica A {\bf 192,} 231 (1993).
\bibitem{015}
O. Derzhko,
cond--mat/9809018.
\bibitem{016}
H. Nishimori,
Phys. Lett. A {\bf 100,} 239 (1984).
\bibitem{017}
O. Derzhko and J. Richter,
Phys. Rev. B {\bf 55,} 14298 (1997);\\
O. Derzhko and J. Richter,
Phys. Rev. B {\bf 59,} 100 (1999).
\bibitem{018}
J. Villain, R. Bidaux, J.-P. Carton, and R. Conte,
J. Physique {\bf 41,} 1263 (1980);\\
E. F. Shender,
Sov. Phys. JETP {\bf 56,} 178 (1982);\\
J. Richter, S. E. Kr\"{u}ger, A. Voigt, and C. Gros,
Europhys. Lett. {\bf 28,} 363 (1994).
\bibitem{019}
O. Derzhko, J. Richter, and O. Zaburannyi,
in preparation.
\end{thebibliography}
\end{document}